\documentclass[a4paper,11pt]{article}
\usepackage{pos}

\title{Photon momentum anisotropies from the late stages of relativistic heavy-ion collisions}
\ShortTitle{Photons from hadronic transport}

\author*[a,b,c,d]{Hannah Elfner}
\author[b,c]{Niklas G\"otz}
\author[e]{Oscar Garcia-Montero}
\author[f,g]{Jean-Francois Paquet}
\author[h]{Charles Gale}

\affiliation[a]{GSI Helmholtzzentrum f\"ur Schwerionenforschung,\\ Planckstra{\ss}e 1, 64291 Darmstadt, Germany}

\affiliation[b]{Institut f\"ur Theoretische Physik, Johann Wolfgang Goethe-Universit\"at,\\ Max-von-Laue-Stra{\ss}e 1, 60438 Frankfurt am Main, Germany}

\affiliation[c]{Frankfurt Institute for Advanced Studies (FIAS), \\ Ruth-Moufang-Stra{\ss}e 1, 60438 Frankfurt am Main}

\affiliation[d]{Helmholtz Research Academy Hesse for FAIR (HFHF), GSI Helmholtz Center,
Campus Frankfurt, \\ Max-von-Laue-Stra{\ss}e 12, 60438 Frankfurt am Main, Germany}

\affiliation[e]{Bielefeld}

\affiliation[f]{Department of Physics and Astronomy, Vanderbilt University, \\ Nashville TN 37240, US}

\affiliation[g]{Department of Mathematics, Vanderbilt University, \\Nashville TN 37240, US}

\affiliation[h]{Department of Physics, McGill University, 3600 University Street, Montreal, QC, H3A 2T8, Canada}

\emailAdd{elfner@itp.uni-frankfurt.de}
\emailAdd{goetz@fias.uni-frankfurt.de}
\emailAdd{garcia@physik.uni-bielefeld.de}
\emailAdd{jean-francois.paquet@Vanderbilt.Edu}
\emailAdd{gale@physics.mcgill.ca}

\abstract{The photon emission from the late stages of the dynamical evolution of heavy-ion reactions at the highest RHIC and LHC energies is investigated. A comparison between a calculation from hadronic rates from a fluid dynamic evolution down to temperatures of 120 MeV and a full non-equilibrium hadronic transport approach is performed. The photon yields are very similar in both cases while the elliptic flow is slightly smaller in the non-equilibrium scenario. This study is important, since it is crucial to apply the same dynamical evolution model for hadronic and electromagnetic observables.}

\FullConference{HardProbes2023\\
 26-31 March 2023\\
 Aschaffenburg, Germany\\}

\begin{document}
\maketitle

\section{Introduction}
Heavy-ion collisions allow to explore the properties of matter under extreme conditions of temperature and density. Since they are highly explosive processes, only the final fragments can be  observed in the detector. Therefore, detailed dynamical modeling of the evolution is crucial to gain insights about the fireball and its properties. Within the last 10 years the consensus has emerged that hybrid approaches combining viscous hydrodynamics for the hot and dense stage and hadronic transport for the late stage of the evolution offer a very good description of hadronic observables at high RHIC and LHC energies (see refs. \cite{Schenke:2020mbo, Petersen:2014yqa} and references therein). 
Photons as electromagnetic probes are particularly interesting since they are emitted from all stages of the reaction and reach the detector undisturbed. The photons of interest here are the direct photons, meaning those that retain after subtracting the decay photons from the total photon yield. The PHENIX collaboration has measured yields and elliptic flow of direct photons that pose a challenge to theoretical descriptions \cite{Adare:2011zr,Adare:2014fwh, Adare:2015lcd}. In this work, we explore how much the late stage non-equilibrium evolution contributes to the yield and the elliptic flow of photons at high beam energies.

\section{Theoretical Framework}

\begin{figure}[h!]
  \includegraphics[width=0.6\textwidth]{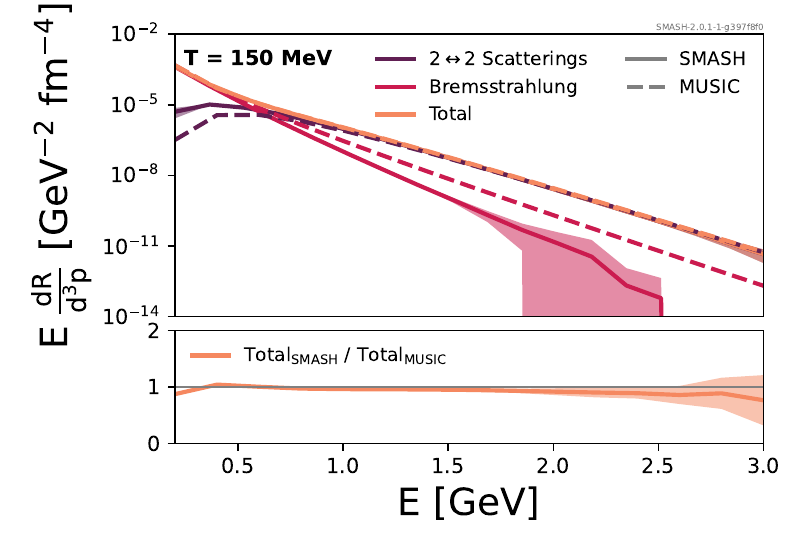}
  \centering
  \caption{Equilibrium photon rate at $T=150$ MeV as calculated in Refs. ~\cite{Turbide:2003si,Liu:2007zzw,Heffernan:2014mla} (dashed lines) compared to the photon rate from an infinite hadronic medium calculated with SMASH (solid lines) for the different contributions. The ratio of the total rate is shown in the lower panel.}
  \label{fig:thermal_rate}
\end{figure}

Instead of a full 3+1 dimensional event-by-event hybrid approach, we restrict ourselves here to a simplified scenario that allows for a quantitative comparison of the equilibrium versus non-equilibrium emission of photons during the late hadronic stage of the reaction. The averaged initial conditions are obtained at $\tau=0.4$ fm/c from the T\textsubscript{R}ENT\textsubscript{O} model \cite{Moreland:2014oya} without transverse flow. The ideal averaged 2+1 dimensional hydrodynamic evolution is performed employing the MUSIC code \cite{Schenke:2010nt,Schenke:2010rr,Paquet:2015lta, MUSIC_link}. The equation of state matches the lattice QCD results and has the same hadronic degrees of freedom as SMASH (Simulating Many Accelerated Strongly-interacting Hadrons) \cite{Weil:2016zrk, SMASH_doi, SMASH_github}, which is used for the hadronic non-equilibrium evolution \cite{Bazavov:2014pvz, Bernhard:2018hnz, eos_code}. The particlization hypersurface is set at a temperature of 150 MeV. Even though this approach is simplified considerably compared to state-of-the-art simulations, the hadronic yields and elliptic flow of pions, kaons and protons are described reasonably well.

For the photon emission from the hadronic stage, we consider two options. In the first case, the photons are produced from the hydrodynamic evolution folded with thermal rates and run down to temperatures of 140 or 120 MeV. The other option is to produce the photons microscopically within the hadronic transport approach. For this purpose the photon production from mesonic $2\leftrightarrow 2$ scatterings and meson bremsstrahlung are taken into account. Please consult \cite{Schafer:2019edr,Schafer:2021slz} for more details. 

Figure \ref{fig:thermal_rate} shows as a benchmark the result of the SMASH calculation in an equilibrated box compared to the equilibrium rate at a temperature of 150 MeV. The individual contributions from $2 \leftrightarrow 2$ scatterings as well as Bremsstrahlung agree very well with each other in the regions where they dominate. The Bremsstrahlung is important at low transverse momentum, while the scatterings contribute at high transverse momentum. This agreement of the thermal rates employed within the hydrodynamic calculation and the microscopic calculation - relying on the individual cross-sections that have been evaluated based on the same effective field theory \cite{Turbide:2003si} - is crucial for the realistic comparison of both scenarios in the dynamic setting that we pursue here. 

\footnote{While these proceedings are based on \cite{Schafer:2021slz}, there was an error in the calculation of the elliptic flow results. This has been corrected now and the figures shown in the following depict the updated results with adapted conclusions.}

\section{Photon Yields and Elliptic Flow}

\begin{figure}[h]
  \includegraphics[width=0.6\textwidth]{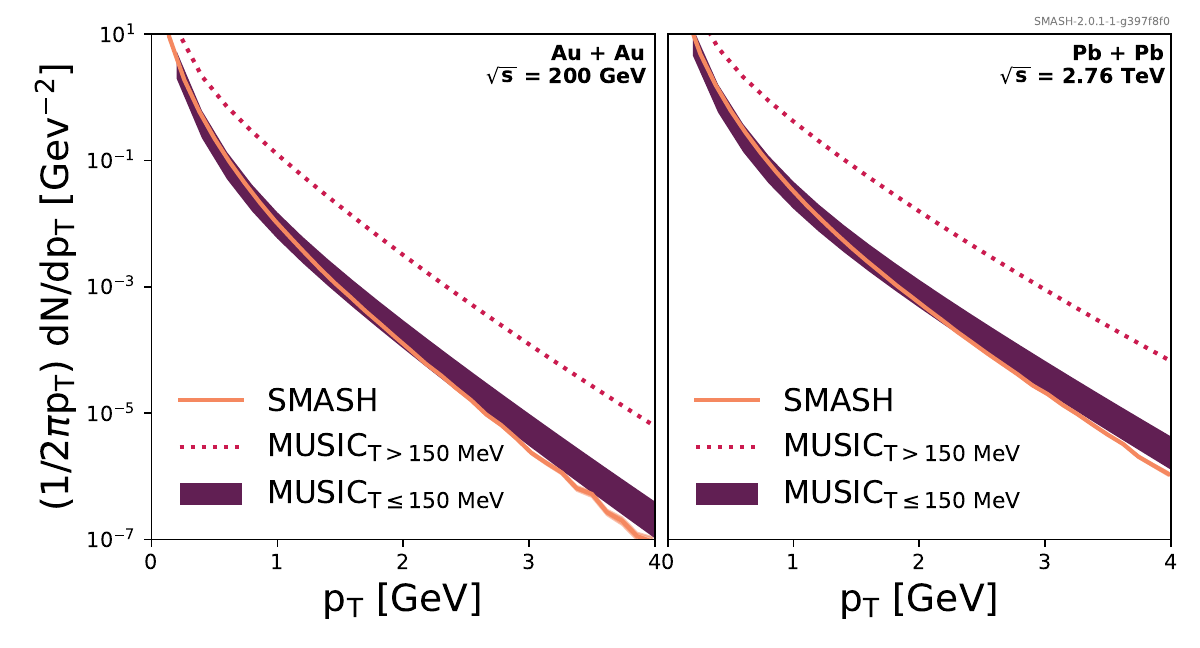}
  \centering
  \caption{Photon yields as a function of transverse momentum for Au+Au collisions at at $\sqrt{\mathrm{s}_\mathrm{NN}}$ = 200 GeV (left) and Pb+Pb collisions at $\sqrt{\mathrm{s}_\mathrm{NN}}$ = 2.76 TeV (right) for a centrality bin of roughly 10-20\% (impact parameter $b=5$fm). The results from non-equilibrium SMASH are depicted with full lines, while the band indicates the equilibrium emission from MUSIC (upper end being emission down to 120 MeV and lower to 140 MeV). The dotted line shows the thermal emission from the evolution before above $T=150$ MeV. }
  \label{fig:pTspectra}
\end{figure}

Within our simplified hybrid calculation, the photon yields are first compared. As shown in Fig. \ref{fig:pTspectra} the yield is dominated by the deconfined evolution above temperatures of 150 MeV. The photon emission as a function of transverse momentum is very similar in the equilibrium and the non-equilibrium scenario, since the line from SMASH overlaps everywhere with the band from the hydrodynamic thermal rates. Having a closer look, the spectrum from SMASH is slightly softer than the one from the equilibrium scenario.

\begin{figure}[t]
  \includegraphics[width=0.99\textwidth]{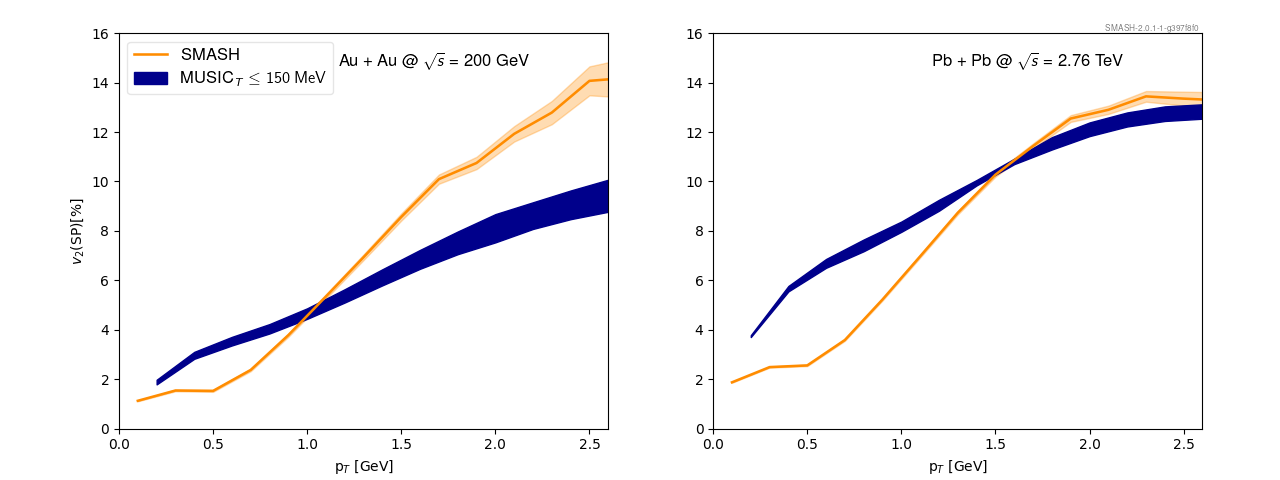}
  \caption{Elliptic flow of photons as a function of transverse momentum for Au+Au collisions at at $\sqrt{\mathrm{s}_\mathrm{NN}}$ = 200 GeV (left) and Pb+Pb collisions at $\sqrt{\mathrm{s}_\mathrm{NN}}$ = 2.76 TeV (right) for a centrality bin of roughly 10-20\% (impact parameter $b=5$fm). Compared is again the equilibrium emission (MUSIC) to the non-equilibrium one from SMASH below 150 MeV temperature.}
  \label{fig:flow_comparison}
\end{figure}

The elliptic flow of photons has been evaluated with the scalar product method \cite{Paquet:2015lta} and referring to the event plane defined by the hadrons as it is done in the experimental analysis. Figure \ref{fig:flow_comparison} shows the comparison of the anisotropic flow from the thermal rate emission (MUSIC) compared with the microscopic non-equilibrium scenario. The contributions are of similar magnitude and at higher transverse momenta at LHC they even overlap completely. Multiple effects contribute to the differences in the elliptic flow. While the asymmetry of the hydrodynamic flow velocity is imprinted on hadrons at particlization, the subsequent anisotropic flow is modified by the larger viscosity in the non-equilibrium evolution. At low transverse momenta the flow is reduced in the microscopic setup while at higher transverse momentum, the particles do not rescatter much and therefore the higher flow values from the hypersurface are preserved. 

\begin{figure}[t]
  \includegraphics[width=0.99\textwidth]{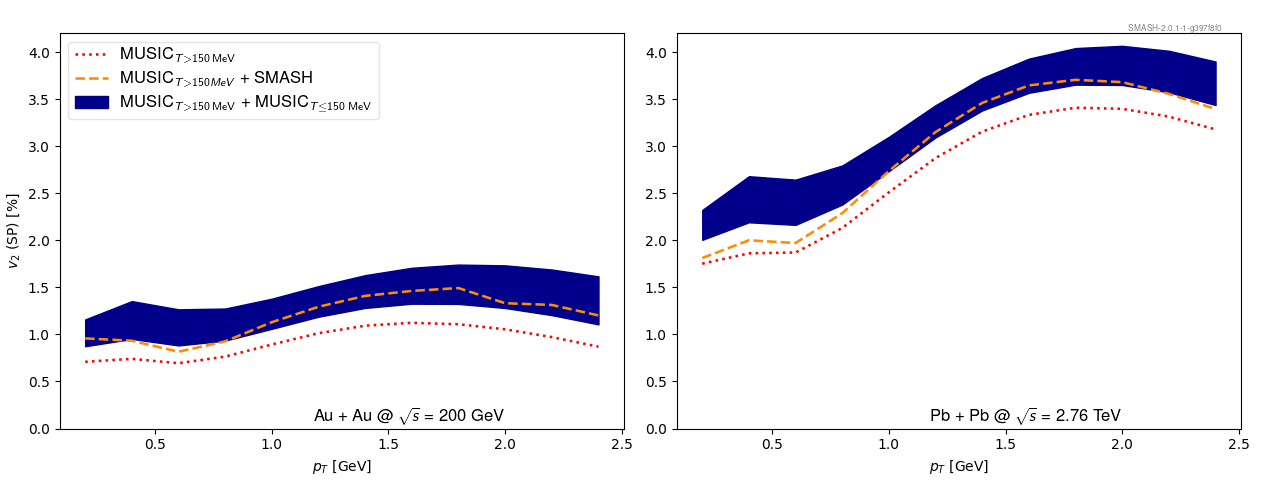}
  \caption{Full result for the total elliptic flow from the hybrid approach. The band indicates the calculation where the photon emission is calculated with MUSIC thermal rates below 150 MeV in temperature (again the upper end corresponds to 120 MeV lowest temperature and the lower end to 140 MeV). The dashed line depicts the calculation invoking SMASH for the photon emission below 150 MeV. The dotted line shows the elliptic flow that has developed in the earlier stages of the evolution.}
  \label{fig:full_flow}
\end{figure}

Figure \ref{fig:full_flow} shows the full result for the elliptic flow for both scenarios. Comparing the band (equilibrium scenario) and the dashed line (non-equilibrium scenario) to the dotted line, one can conclude that a significant portion of elliptic flow is developed in the late hadronic stage of the evolution. From a transverse momentum of ~1 GeV the calculations from MUSIC and SMASH agree very well. At low transverse momentum the higher viscosity in the non-equilibrium hadronic transport approach leads to a reduced elliptic flow compared to the equilibrium calculation. The differences become only relevant at very low transverse momentum where currently no experimental measurements exist.

\section{Conclusions}
As high energy heavy-ion physics is approaching an era of precision measurements and detailed theoretical calculations, it is important to describe many observables within the same approach. Therefore, we have employed a hybrid approach and compared the photon yields and elliptic flow from the final stage hadronic stage. The more realistic microscopic non-equilibrium hadronic transport photon emission only differs significantly from the hydrodynamic estimate at very low transverse momenta (<1 GeV) and leads to a slight reduction of elliptic flow while the yields are very similar. It would be interesting to employ the SMASH hadronic afterburner with consistent photon emission in a realistic event-by-event hybrid approach and look at more observables including $v_3$ and perform a comparison to experimental results.

\section{Acknowledgements}
This project was supported in part by the DAAD funded by BMBF with Project-ID 57314610, in part by the Deutsche Forschungsgemeinschaft (DFG, German Research Foundation) – Project number 315477589 – TRR 211, and in part by the Natural Sciences and Engineering Research Council of Canada. Computational resources have been provided by the Center for Scientific Computing (CSC) at the Goethe-University of Frankfurt and the GreenCube at GSI. Computations were also made on the supercomputer B\'{e}luga, managed by Calcul Qu\'{e}bec and by the Digital Research Alliance of Canada. The operation of this supercomputer is funded by the Canada Foundation for Innovation (CFI), Minist\`{e}re de l'\'{e}conomie et de l'Innovation du Qu\'{e}bec (MEI) and le Fonds de recherche du Qu\'{e}bec (FRQ).

\end{document}